\documentclass[lettersize,journal]{IEEEtran}
\usepackage{amsmath,amsfonts}
\usepackage{algorithmic}
\usepackage{algorithm}
\usepackage{array}
\usepackage[caption=false,font=normalsize,labelfont=sf,textfont=sf]{subfig}
\usepackage{textcomp}
\usepackage{stfloats}
\usepackage{url}
\usepackage{verbatim}
\usepackage{graphicx}
\usepackage{cite}
\usepackage{multirow}
\usepackage{url}
\usepackage{subfig}
\usepackage{graphicx}
\usepackage{bm}
\usepackage{array}    
\usepackage{makecell} 
\usepackage{booktabs} 
\usepackage{diagbox}
\usepackage{tabularx} 
\usepackage{subfig}          
\usepackage{caption}         
\usepackage{color}
\begin{document}

\title{Tiny-WiFo: A Lightweight Wireless Foundation Model for Channel Prediction via Multi-Component Adaptive Knowledge Distillation}

\author{Haotian Zhang,~\IEEEmembership{Graduate Student Member,~IEEE}, Shijian Gao,~\IEEEmembership{Member,~IEEE}, Xiang Cheng,~\IEEEmembership{Fellow,~IEEE}

\thanks{Haotian Zhang and Xiang Cheng are with the State Key Laboratory of Photonics and Communications, School of Electronics, Peking University, Beijing 100871, P. R. China (e-mail: haotianzhang@stu.pku.edu.cn; xiangcheng@pku.edu.cn).
			
Shijian Gao is with the Internet of Things Thrust, The Hong Kong University of Science and Technology (Guangzhou), Guangzhou 511400, P. R. China (e-mail: shijiangao@hkust-gz.edu.cn).

}}

\markboth{Journal of \LaTeX\ Class Files,~Vol.~14, No.~8, August~2021}%
{Shell \MakeLowercase{\textit{et al.}}: A Sample Article Using IEEEtran.cls for IEEE Journals}


\maketitle

\begin{abstract}
The massive scale of Wireless Foundation Models (FMs) hinders their real-time deployment on edge devices. This letter moves beyond standard knowledge distillation by introducing a novel Multi-Component Adaptive Knowledge Distillation (MCAKD) framework. Key innovations include a Cross-Attention-Based Knowledge Selection (CA-KS) module that selectively identifies critical features from the teacher model, and an Autonomous Learning-Passive Learning (AL-PL) strategy that balances knowledge transfer with independent learning to achieve high training efficiency at a manageable computational cost. When applied to the WiFo FM, the distilled Tiny-WiFo model, with only 5.5M parameters, achieves a 1.6 ms inference time while retaining over 98\% of WiFo’s performance and its crucial zero-shot generalization capability, making real-time FM deployment viable.
\end{abstract}


\vspace{-1.2em}
\section{Introduction}
\IEEEPARstart{W}{i}reless foundation models (FM) pre-trained on large-scale heterogeneous datasets have emerged as a new paradigm for wireless channel-related tasks \cite{wifo,wirelessllm,musefm}.  However, their practical deployment is hindered by large parameter sizes and slow inference speeds. In dynamic scenarios, inference latency can exceed the channel coherence time, rendering predictions obsolete \cite{wifo,Dai_Trans}. Furthermore, the massive model size makes fine-tuning for new environments prohibitively expensive. These limitations critically challenge their use in real-time applications and edge device deployment.

To mitigate these issues, model compression techniques, such as pruning, quantization, and knowledge distillation (KD), have been studied. Among them, KD \cite{kd2} has shown great potential in transferring knowledge from a large teacher model to a compact student model, effectively maintaining performance while reducing model size and latency. Unlike pruning and quantization, which are often constrained by hardware support for sparse or low-precision computation, KD produces a dense and hardware-agnostic student model, facilitating deployment.
Recently, KD has been widely explored for lightweighting large language models (LLM), focusing on aligning the student's output distribution with the teacher's. For example,  \cite{MiniLLM} employed the reverse KLD to improve model's long-text generation ability. In \cite{CLIP}, a targeted distillation method was proposed with token selection and confidence weighting to enhance low-shot and domain-shifted tasks. 
In the field of wireless communications, Guo {\em et al}. \cite{feedback_dis} leveraged KD to obtain a lightweight proxy channel state information (CSI) decoder. However, the approach primarily focuses on prediction consistency, neglecting structural knowledge. Similarly, \cite{CLST} introduces signal knowledge distillation yet remains centered on soft labels, failing to fully utilize intermediate representations. In summary, most existing KD schemes designed for LLMs or wireless communication models remain confined to mimicking output distributions, overlooking the importance of task-oriented internal knowledge extraction. This limitation hinders the preservation of key knowledge crucial to FM’s performance and generalization capability. 

To bridge this gap, we propose a Multi-Component Adaptive Knowledge Distillation (MCAKD) scheme, tailored for transformer-based wireless foundation models. MCAKD scheme distills the knowledge embedded in teacher's CSI embedding module, along with the attention weights and hidden states of its encoder and decoder modules, into the student model. To enable selective knowledge transfer while handling the dimensional mismatch, we design a Cross-Attention-Based Knowledge Selection (CA-KS) module that selects the most relevant knowledge from the teacher. Furthermore, we introduce a computationally efficient Autonomous Learning-Passive Learning (AL-PL) training strategy. It strategically alternates between self-supervised learning and distillation phases. This approach reduces the overhead of frequently querying the large teacher model while effectively balancing knowledge imitation with the student's self-discovery.

We validate the effectiveness of MCAKD using WiFo \cite{wifo}, a recently proposed channel prediction FM renowned for its exceptional performance and zero-shot generalization. Applying our method yields Tiny-WiFo, a highly efficient variant that that retains over $98\%$ of WiFo's accuracy with only $5.5$M parameters and $1.6$ ms inference time. Critically, Tiny-WiFo not only maintains the teacher's zero-shot generalization but even surpasses it on unseen scenarios.

\vspace{-1em}
\section{System Model and Problem Description}
\subsection{System Model}
We consider a multiple-input single-output orthogonal frequency-division multiplexing (MISO-OFDM) system where a base station employing a $N_v \times N_h$ uniform planar array  serves a single-antenna user. $N_v$ and $N_h$ represent the number of vertical and horizontal antennas, with $N=N_v \times N_h$. 
The considered time-frequency region spans $T$ and $K$ resource blocks (RB) along the time and frequency dimensions, respectively. 
The pilot placement pattern is uniform across all antennas, with each RB containing one pilot. The spatial-temporal-frequency (STF) CSI is represented as $\mathbf{H}\in\mathbb{C}^{T\times K \times N}$. We only consider the channel at the pilot locations within each RB, meaning that all elements of $\mathbf{H}$ are known channel estimates.
\vspace{-1.5em}
\subsection{Problem Description}
This work aims at crafting a lightweight foundational model to deal with two types of channel prediction tasks:
\begin{itemize}
\item[$\bullet$] \textbf{Time-domain prediction}: Forecast the future $(T-X_{\rm T})$ RBs from the historical $X_{\rm T}$ RBs. Denoting a time-domain channel prediction model as $f_{\rm T}$ and its output as $\hat{\mathbf{H}}$, the process is:
$\hat{\mathbf{H}}[X_{\mathrm{T}}+1:T,:,:]=f_{\mathrm{T}}(\mathbf{H}[1:X_{\mathrm{T}},:,:])$.

\item[$\bullet$] \textbf{Frequency-domain prediction}: Infer the subsequent $(K-X_{\rm F})$ RBs from the first $X_{\rm F}$ RBs. Denoting a frequency-domain channel prediction model as $f_{\rm F}$, the process is:
$\hat{\mathbf{H}}[:,X_{\mathrm{F}}+1:K,:]=f_{\mathrm{F}}(\mathbf{H}[:,1:X_{\mathrm{F}},:])$.

\end{itemize}

\vspace{-0.4em}
\section{Tiny-WiFo: A lightweight channel prediction foundation model}
\begin{figure*}
    \centering
    \includegraphics[width=0.99\linewidth]{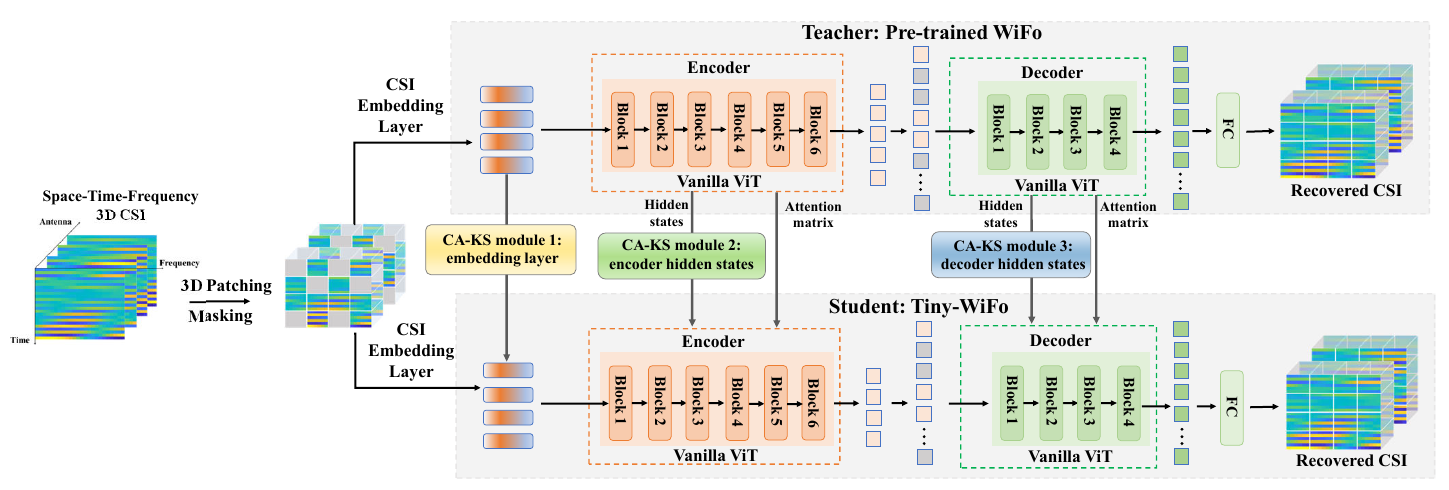}
    \caption{An overview of MCAKD scheme. }
    \label{fig:enter-label}
    \vspace{-1.1em}
\end{figure*}

\subsection{Preliminaries}
WiFo is the first foundational model designed to achieve time-frequency channel prediction. It is initially pretrained on a massive and diverse set of CSI datasets and can be directly applied to channel prediction task in any communication system configuration without additional fine-tuning. 

As shown in Fig.~\ref{fig:enter-label}, WiFo utilizes a transformer-based encoder-decoder architecture, built from four key components: 3D masking, CSI embedding, an encoder, and a decoder. It frames time- and frequency-domain channel prediction as a unified reconstruction problem.  To model the intricate STF correlations inherent in CSI, three self-supervised training tasks are adopted: time-domain/frequency-domain/random masked reconstruction. In random-masked reconstruction, tokens from STF dimensions are randomly masked with a $R_r$ ratio. In time- and frequency-domain masked reconstruction, tokens are masked along the corresponding dimensions with ratios $R_t$ and $R_f$.

During the pretraining phase, the masking operations are performed to let WiFo learn the inherent STF correlations within CSI. In the inference phase, masking strategies are applied to perform specific channel prediction tasks. For example, in time-domain channel prediction, WiFo masks CSI after the current time instant $t$ and predicts it.

\vspace{-0.9em}
\subsection{Proposed MCAKD scheme}
As shown in Fig.~\ref{fig:enter-label}, the architecture of Tiny-WiFo remains generally same to that of the teacher WiFo, with the only difference being that the hidden size dimension is reduced to half, while the depth remains unchanged. Given this structural similarity, we propose a per-layer MCAKD scheme for Tiny-WiFo, which consists of the following three components:

\subsubsection{Attention weights distillation}
The attention weights learned by WiFo’s transformer encoder and decoder reflect the 3D correlations among antenna elements, time slots, and subcarriers. This demonstrates how the model focuses on different parts of CSI to reconstruct missing information. These attention patterns encode physical properties such as multi-path propagation and Doppler shifts, which are fundamental to accurate channel prediction.

Unlike common approaches that use mean squared error (MSE) to enforce exact value matching between teacher and student, we employ a cosine loss for attention distillation. This is because cosine loss enforces directional alignment between the teacher’s and student’s intermediate results while ignoring their absolute magnitudes. Such scale-invariant guidance gives the student greater optimization freedom, letting it follow the teacher’s cue without being forced to replicate exact values. Let $\mathbf{A}_{\rm t}^{\rm e} \in \mathbb{R}^{L^{e} \times N_{\rm h}\times L\times L}$ and $\mathbf{A}_{\rm s}^{\rm e} \in \mathbb{R}^{L^{e} \times N_{\rm h}\times L\times L}$ be the encoder attention weight matrix of teacher WiFo and Tiny-WiFo, where $L^e$, $N_{\rm h}$, and $L=T\times N$ denote the depth of encoder, number of heads, and the sequence length, respectively. Similarly, let $\mathbf{A}_{\rm t}^{\rm d} \in \mathbb{R}^{L^{d} \times N_{\rm h}\times 2L\times 2L}$ and $\mathbf{A}_{\rm s}^{\rm d} \in \mathbb{R}^{L^{d} \times N_{\rm h}\times 2L\times 2L}$ be the decoder attention weight matrix of teacher model and Tiny-WiFo, where $L^d$ denotes the depth of decoder. For the 
$i$-th head in the $l$-th layer, we first flatten the attention matrix by matricising the last two dimensions, i.e., $\tilde{\mathbf{A}}_{\rm t}^{\rm e} = \rm{flat}(\mathbf{A}_{\rm t}^{\rm e})  \in \mathbb{R}^{L^{e} \times N_{\rm h}\times L^2},\tilde{\mathbf{A}}_{\rm t}^{\rm d} = \rm{flat}(\mathbf{A}_{\rm t}^{\rm d})  \in \mathbb{R}^{L^{d} \times N_{\rm h}\times 4L^2}$. 

{\small
{\setlength{\abovedisplayskip}{1pt}%
\setlength{\belowdisplayskip}{1pt}%
\begin{equation}
\mathcal{L}_{\text{attn}}
= \sum_{m\in\{\rm{e},\rm{d}\}}
1-
\frac{1}{N_{\rm h} L^{m}}
\sum_{i=1}^{N_{\rm h}}
\sum_{l=1}^{L^{m}}
\textit{CosSim}\bigl(
\tilde{\mathbf{A}}^{m}_{\rm t}[l,i,:],
\tilde{\mathbf{A}}^{m}_{\rm s}[l,i,:]
\bigr).
\end{equation}}}where $\textit{CosSim}(\mathbf{x}_1,\mathbf{x}_2)=\frac{\mathbf{x}_1\cdot \mathbf{x}_2}{\Vert\mathbf{x}_1\Vert_2\Vert\mathbf{x}_2\Vert_2}$ represents the cosine similarity between $\mathbf{x}_1$ and $\mathbf{x}_2$.

\subsubsection{Embedding-layer distillation}
The CSI embedding module plays a critical role in converting CSI data into a tokenized format suitable for transformer processing. This embedding captures essential structural priors and invariance properties that are crucial for handling variable CSI dimensions.

A fundamental challenge in distilling knowledge from the teacher WiFo to Tiny-WiFo is the dimensionality mismatch in their hidden states. Teacher WiFo's CSI tokens have a higher dimensionality than Tiny-WiFo's, which prevents the direct computation of a similarity loss between their raw CSI tokens. Moreover, directly distilling all dimensions from the teacher to the student is inefficient, as not all teacher knowledge is equally relevant to the student. To overcome this, we design the CA-KS module\footnote{\color{black}The CA‑KS operates at the feature‑representation level and selects knowledge based on cross‑model relevance rather than the semantic content of the features, making it inherently mask‑agnostic and functionality-agnostic. Hence, we instantiate three separate CA-KS modules, one each for the CSI embedding module, encoder, and decoder. The same set of modules is shared across three masking strategies.}. Let $\mathbf{E}_{\rm t} \in \mathbb{R}^{2L \times D_{\rm t}}$ and $\mathbf{E}_{\rm s} \in \mathbb{R}^{2L \times D_{\rm s}}$ denote the teacher WiFo's CSI tokens and Tiny-WiFo's CSI tokens obtained from their CSI embedding modules, where $D_{\rm t}$ and $D_{\rm s}$ represent teacher WiFo's hidden dimension and Tiny-WiFo's hidden dimension, respectively. The working process of the CA-KS module is outlined step by step below:

\textbf{Step 1: Cross-Attention Based Relevance Computation.} To identify which teacher dimensions are most relevant to the student's representation, the CA-KS module employs a cross-attention mechanism to compute the relevance of each dimension in the teacher WiFo's CSI tokens  $\mathbf{E}_{\rm t}$ with respect to Tiny-WiFo's CSI tokens $\mathbf{E}_{\rm s}$. Tiny-WiFo's CSI tokens serve as the Query, while the teacher WiFo's CSI tokens are the Key and Value. They are projected by learnable matrices: $\mathbf{Q}=\mathbf{E}_{\rm s}\mathbf{W}_{\rm q}$, $\mathbf{K}=\mathbf{E}_{\rm t}\mathbf{W}_{\rm k}$, $\mathbf{V}=\mathbf{E}_{\rm t}\mathbf{W}_{\rm v}$, where $\mathbf{W}_{\rm q}$, $\mathbf{W}_{\rm k}$, and $\mathbf{W}_{\rm v}$ represent the projection matrices. Afterwards, multi-head attention is applied to compute attention weights:
{
\setlength{\abovedisplayskip}{6pt}%
  \setlength{\belowdisplayskip}{6pt}%
\begin{equation}
\mathbf{O}_{\rm attn}, \mathbf{A}_{\rm attn} = \rm{MultiHeadAttention}(\mathbf{Q},\mathbf{K},\mathbf{V}),
\end{equation}}
where $\mathbf{A}_{\rm attn} \in \mathbb{R}^{2L \times 2L}$ is the attention weight matrix, representing the similarity between Tiny-WiFo's and teacher's CSI tokens. $\mathbf{O}_{\text{attn}} \in \mathbb{R}^{2L \times D_{\rm t}}$ is the weighted sum output.

 \textbf{Step 2: Importance Scoring \& Ranking.} The importance score $\mathbf{s} \in \mathbb{R}^{1\times D_{\rm t}}$ per teacher dimension is obtained by summing over sequence dimensions with the attention weights:
{
\setlength{\abovedisplayskip}{6pt}%
  \setlength{\belowdisplayskip}{6pt}%
\begin{equation}
\mathbf{s}(d) = \sum_{l=1}^{2L}\sum_{s=1}^{2L}\mathbf{A}_{\rm attn}[l,s] \mathbf{E}_{\rm t}[s,d], \quad d=1,2,\cdots,D_{\rm t}.
\end{equation}
}
We then sort the importance score $\mathbf{s}$ in descending order and obtain the sorted indices: $\mathbf{I} = \mathop{\mathrm{argsort}}(\mathbf{s},\text{descending})$.
The top $D_{\rm s}$ most important indices are selected: $\mathbf{I}_{D{\rm s}}=\mathbf{I}(1:D_{\rm s})$.

\textbf{Step 3: Knowledge Selection.} Finally, leveraging the ranked indices, we perform a selective transfer of knowledge. For each sequence position $l \in {1,\cdots,2L}$, the selection operation retrieves $D_{\rm s}$ feature values from the teacher's CSI tokens vector $\mathbf{E}_{\rm t}[l,:]$ using the index set $\mathbf{I}_{D_{\rm s}}$:
{
\setlength{\abovedisplayskip}{6pt}%
  \setlength{\belowdisplayskip}{6pt}%
\begin{equation}
\tilde{\mathbf{E}}_{\rm t}[l,j] = \mathbf{E}_{\rm t}[l,\mathbf{I}_{D_{\rm s}}[j]], \quad j = 1,2,\cdots, D_s,
\end{equation}}
where $\tilde{\mathbf{E}}_{\rm t} \in \mathbb{R}^{2L \times D_{\rm s}}$ represents the filtered CSI tokens.

After filtering the teacher WiFo's CSI tokens knowledge using the CA-KS module, the embedding layer distillation loss is calculated by:
{
\setlength{\abovedisplayskip}{1pt}%
\setlength{\belowdisplayskip}{1pt}%
\begin{align}
\mathcal{L}_{\rm embed} &= 1-\frac{1}{2L}\sum_{l=1}^{2L}\textit{CosSim}(\tilde{\mathbf{E}}_{\rm t}[l,:],\mathbf{E}_{\rm s}[l,:]). \label{eq:emb_loss}
\end{align}
}

\subsubsection{Hidden states distillation}
Similar to attention weights, the hidden states produced by WiFo’s encoder and decoder layers contain high-dimensional representations that encapsulate both local and global STF features of the channel. These representations are optimized through large-scale pre-training to generalize across diverse configurations and scenarios.

The distillation process of embedding-layer is applied in an identical manner to the hidden states taken after the last multi-head attention sub-layer of both the encoder and decoder. Let $\tilde{\mathbf{M}}^{\rm e}_{\rm t} \in \mathbb{R}^{L\times D_{\rm s}}$ and ${\mathbf{M}}^{\rm e}_{\rm s} \in \mathbb{R}^{L\times D_{\rm s}}$ denote the encoder hidden-state matrix of the teacher WiFo filtered by CA-KS module and the encoder hidden-state matrix of Tiny-WiFo, respectively. Similarly, let $\tilde{\mathbf{M}}^{\rm d}_{\rm t} \in \mathbb{R}^{2L\times D_{\rm s}}$ and ${\mathbf{M}}^{\rm d}_{\rm s} \in \mathbb{R}^{2L\times D_{\rm s}}$ be the decoder hidden-state matrix of the teacher WiFo filtered by CA-KS module and  the decoder hidden-state matrix of Tiny-WiFo, respectively. Let $L^{\rm{e}}=L$ and $L^{\rm{d}}=2L$. With this notation, the hidden states distillation loss is defined as:
{
\begin{equation}
\mathcal{L}_{\text{hs}}
= \sum_{m\in\{\rm{e},\rm{d}\}}
1-
\frac{1}{L^{m}}
\sum_{l=1}^{L^{m}}
\textit{CosSim}\bigl(
\tilde{\mathbf{M}}^{m}_{\rm t}[l,:],
\mathbf{M}^{m}_{\rm s}[l,:]
\bigr)
.
\end{equation}}

Finally, the overall distillation loss is formed by equally weighted summation of the above three components:
{
\setlength{\abovedisplayskip}{6pt}%
  \setlength{\belowdisplayskip}{6pt}%
\begin{equation}
    \label{dis_loss}
    \mathcal{L}_{\text{MCAKD}} = \mathcal{L}_{\text{attn}}+\mathcal{L}_{\text{embed}}+\mathcal{L}_{\text{hs}}.
\end{equation}}
In addition to the distillation loss $\mathcal{L}_{\text{MCAKD}}$ used for distilling knowledge from the teacher WiFo, Tiny-WiFo also requires self-supervised training loss to learn how to perform the masked reconstruction task. As shown in Eq.~\eqref{hard_loss}, it is the MSE between the recovered CSI $\hat{\mathbf{H}}[\omega]$ and ground truth CSI ${\mathbf{H}}[\omega]$, where $\omega$ denotes the predicted CSI subset in either the temporal or frequency domain.
\begin{equation}
    \label{hard_loss}
    \mathcal{L}_{\text{mse}} = \frac{1}{|\omega|}\Big\Vert \mathbf{H}[\omega]-\hat{\mathbf{H}}[\omega]\Big\Vert^2_{\rm F}
\end{equation}

\vspace{-1.6em}
\subsection{Efficient Training with AL-PL Strategy}

The training of Tiny-WiFo employs an Autonomous Learning-Passive Learning (AL-PL) strategy, designed to address the significant computational and temporal costs associated with KD. Leveraging the teacher model is computationally intensive, as it requires a full forward pass in every epoch of traditional KD. {\color{black}Even if intermediate teacher outputs are pre-stored to avoid frequent forward passes, the resulting storage overhead would be prohibitively large, making it impractical for real-world implementation.} Our AL-PL strategy mitigates this burden by strategically interspersing costly distillation phases (Passive Learning, $\mathcal{P}_{\rm d}$) with periods dedicated to autonomous self-supervised learning (Autonomous Learning, $\mathcal{P}_{\rm s}$). The transition to passive learning is carefully determined based on the student's autonomous learning progress, ensuring that the teacher's guidance is incorporated at the most beneficial moments. 
This design strikes an optimal balance, preventing over-dependence on the teacher while still leveraging its expertise when most needed.
Consequently, the student model, informed by the direction of the teacher's knowledge, actively constructs its own specialized representations that are better fitted to its structural constraints.

Under the proposed AL-PL strategy, the training objective for Tiny-WiFo alternates between different learning phases. Specifically, in the $e$-th training epoch, the training loss for Tiny-WiFo is determined by the learning phases:
{
\setlength{\abovedisplayskip}{3pt}%
  \setlength{\belowdisplayskip}{3pt}%
\begin{equation}
    \label{indicator}
    \mathcal{L}=\left\{
    \begin{array}{cl}
        \mathcal{L}_{\text{mse}},  &  e\in \mathcal{P}_{\rm s}   \\
        \mathcal{L}_{\text{mse}}+ \lambda\mathcal{L}_{\text{MCAKD}},  &  e\in \mathcal{P}_{\rm d}
    \end{array} \right.,
\end{equation}}
where $\mathcal{P}_{\rm s}$ and $ \mathcal{P}_{\rm d}$ represent the autonomous learning phase and the passive learning phase, respectively. 

\vspace{-0.4em}
\section{Experimental Results}
\subsection{Implementation Details}
\subsubsection{Baselines}
To thoroughly evaluate the MCAKD scheme as well as the Tiny-WiFo model, we compare them with the following groups of models:
\begin{itemize}
\item \textbf{Teacher and Non-Distilled Model}: The teacher {WiFo} model and {\color{black}Tiny-WiFo w/o KD} are compared, where the latter shares the same architecture as Tiny-WiFo but is pre-trained without knowledge distillation.
\item \textbf{Channel Prediction Models}:
\begin{enumerate}
\item \textbf{State-of-the-Art Schemes}: These include the {Transformer-based} prediction model \cite{Dai_Trans}, and {LLM4CP*} scheme \cite{llm4cp}, which is fine-tuned from GPT-2 to achieve channel prediction. 
\item \textbf{Classical Model}: A {two-layer long short-term memory (LSTM)} based prediction scheme \cite{lstm}.
\end{enumerate}
\item \textbf{Alternative Distilled Model}: {\color{black}We additionally built another lightweight model obtained by applying a classical logits-matching KD method \cite{CLST} to the teacher WiFo, denoted as Tiny-WiFo w/ LMKD}. This model shares the same architecture as Tiny-WiFo, serving as a baseline to benchmark the effectiveness of the  MCAKD scheme.
\end{itemize}

\subsubsection{Datasets}
A total of $18$ datasets are included, covering six 5G New Radio frequency bands, eight 3GPP scenarios, and various user speeds. The D1–D16 datasets are included in the pretraining phase of both teacher WiFo and Tiny-WiFo, while the D17 and D18 datasets are utilized for generalization testing, as their carrier frequencies are not covered in the D1–D16 datasets. Detailed simulation configurations for each dataset can be found in \cite{wifo}. 

\begin{table*}[!t]
\setlength{\tabcolsep}{3pt} %
\renewcommand{\arraystretch}{0.8} 
\centering
\caption{The NMSE performance of Tiny-WiFo and other baselines on the time/frequency-domain channel prediction.}
\label{table_example}
{\scriptsize
\begin{tabular}{c | c | c | c | c | c | c | c}
\toprule
\textbf{\diagbox[width=25mm]{Datasets}{Methods}} & \textbf{Tiny-WiFo} & {\color{black}\textbf{Tiny-WiFo w/o KD}} & {\color{black}\textbf{Tiny-WiFo w/ LMKD}} & \textbf{Teacher  WiFo} & \textbf{Transformer} & \textbf{LSTM} & \textbf{LLM4CP*} \\ 
\midrule
D1--D16 & {\underline{$-6.75$} / \underline{$-7.92$}} & {$-6.56$ / $-7.76$} & {$-6.60$ / $-7.70$} & {\bm{$-6.78$} / \bm{$-8.01$}} & {$-4.88$ / $-4.38$} & {$-2.51$ / $-2.59$} & {$-6.27$ / $-7.01$} \\ 
\midrule
D17 & {\underline{$-5.17$} / \underline{$-6.11$}} & {$-5.05$ / $-6.02$} & {$-5.02$ / $-5.94$} & {\bm{$-5.27$} / \bm{$-6.32$}} & {{$-1.76$} / {$-0.52$}} &{$0.70$ / $-0.12$} & {$-4.60$ / $-4.49$} \\ 
\midrule
D18 & {\bm{$-3.88$} / \bm{$-5.59$}} & {$-3.76$ / $-5.24$} & {$-3.81$ / $-5.39$} & {\underline{$-3.77$} / \underline{$-5.53$}} & {$-1.56$ / $-1.78$} & {$0$ / $-0.01$} & {$-3.72$ / $-4.99$}\\
\bottomrule
\end{tabular}}
\vspace{-1.4em}
\end{table*}

\subsubsection{Hyper-parameters settings}
Both the Tiny-WiFo and the teacher have an encoder depth of $6$, a decoder depth of $4$, and $8$ attention heads. The hidden dimension of Tiny-WiFo is $256$, while that of the teacher WiFo is $512$. The weight for the distillation loss $\lambda$ is set to $0.1$. The ADAM optimizer is used, with a base learning rate of $5\times 10^{-4}$ and a batch size of $64$.
\vspace{-0.8em}
\subsection{Channel Prediction Performance}
We first evaluate the performances of Tiny-WiFo and baselines in Table \ref{table_example}, where the best results are highlighted in bold and the second-best results are underlined. Normalized mean square error (NMSE) is adopted as channel prediction performance metric given by $\text{NMSE}=10\lg(\Vert \hat{\mathbf{H}}-\mathbf{H} \Vert^2_{\rm F}/\Vert \mathbf{H} \Vert^2_{\rm F})$. In Table \ref{table_example}, each cell presents the time-domain NMSE (left) and frequency-domain NMSE (right) separated by a slash.

Tiny-WiFo outperforms the non-distilled baseline Tiny-WiFo w/o KD across all datasets, demonstrating that the MCAKD scheme is effective in improving the performance of small models across diverse system configurations. Moreover, Tiny-WiFo surpasses the Tiny-WiFo w/ LMKD baseline by a clear margin. This performance gap validates the superiority of our MCAKD scheme over widely-used output distribution imitation approach. It is also noteworthy that the Tiny-WiFo w/ LMKD underperforms even the non-distilled model Tiny-WiFo w/o KD. This indicates that an inappropriate distillation strategy can be detrimental and degrade the student's performance. Furthermore, Tiny-WiFo outperforms the state-of-the-art methods by at least $0.16$ dB and up to $5.59$ dB NMSE across all datasets, establishing a clear performance lead. Notably, compared to the lightweight LSTM-based approach, Tiny-WiFo achieves significantly better prediction accuracy and stronger generalization, particularly on unseen scenarios. Across all datasets, the NMSE difference between Tiny-WiFo and its teacher remains within $0.21$ dB, demonstrating that the  Tiny-WiFo achieves nearly lossless compression. Importantly, Tiny-WiFo successfully inherits the excellent generalization capability of the teacher. {\color{black} An interesting observation is that Tiny-WiFo outperforms the teacher on D18. We speculate that the compact model, together with the distilled knowledge, happens to match the characteristics of D18 better. Nevertheless, the teacher still performs better in general cases.}

\vspace{-1.2em}
\subsection{Ablation Study}
In this section, we investigate the impact of different components of the distillation loss, the AP-PL strategy, and the CA-KS module on the performance of Tiny-WiFo. In every cell of Table \ref{tab:ablation}, the upper number is the time-domain NMSE and the lower number is the frequency-domain NMSE. As shown in Table \ref{tab:ablation}, every individual component contributes to the performance, with the AL-PL strategy proving to be crucial. Ablating the AL-PL strategy leads to a significant performance drop, particularly in frequency-domain prediction (e.g., $-7.92$ dB to $-7.55$ dB on D1-D16), underscoring its importance in enabling the student to effectively balance knowledge imitation with self-discovery. Furthermore, the knowledge of attention matrix is the most influential factor for time-domain prediction performance, while the AL-PL strategy is most important for frequency-domain prediction.
\begin{table}[!t] 
\setlength{\tabcolsep}{3.1pt}      
\renewcommand{\arraystretch}{0.8}
\scriptsize                    
\centering
\caption{Ablation studies of Tiny-WiFo.}
\label{tab:ablation}

\newcolumntype{C}{>{\centering\arraybackslash}X}

\begin{tabularx}{\linewidth}{C|C|C|C|C|C|C}
\toprule
\textbf{\makebox[0pt][c]{Datasets}} &
\textbf{\makebox[0pt][c]{Tiny-WiFo}} &
\textbf{\makecell{w/o\\$\mathcal L_{\text{embed}}$}} &
\textbf{\makecell{w/o\\$\mathcal L_{\text{att}}$}} &
\textbf{\makecell{w/o\\$\mathcal L_{\text{hs}}$}} &
\textbf{\makecell{w/o\\CA-KS}} &
\textbf{\makecell{w/o\\AL-PL}} \\
\midrule
D1–D16 &
\makecell{\bm{$-6.75$}\\\bm{$-7.92$}} &
\makecell{$-6.63$\\$-7.78$} &
\makecell{$-6.47$\\$-7.84$} &
\makecell{$-6.67$\\$-7.86$} &
\makecell{$-6.63$\\$-7.89$} &
\makecell{$-6.54$\\$-7.55$} \\
\midrule
D17 &
\makecell{\bm{$-5.17$}\\$-6.11$} &
\makecell{$-5.02$\\$-5.94$} &
\makecell{$-4.93$\\$-6.13$} &
\makecell{$-5.00$\\\bm{$-6.19$}} &
\makecell{$-5.04$\\$-6.13$} &
\makecell{$-5.06$\\$-5.95$} \\
\midrule
D18 &
\makecell{\bm{$-3.88$}\\\bm{$-5.59$}} &
\makecell{$-3.74$\\$-5.43$} &
\makecell{$-3.70$\\$-5.56$} &
\makecell{$-3.76$\\$-5.50$} &
\makecell{$-3.85$\\$-5.48$} &
\makecell{$-3.82$\\$-5.36$} \\
\bottomrule
\end{tabularx}
\vspace{-0.6em}
\end{table}

{\color{black}
\vspace{-1.2em}
\subsection{Tradeoff Between Model Size and Performance}
\begin{table}[!t]
\setlength{\tabcolsep}{1.1pt}     
\renewcommand{\arraystretch}{0.6}
\centering
\caption{{\color{black}Performance-model size trade-off of Tiny-WiFo.}}
{\scriptsize
\begin{tabular}{>{\centering\arraybackslash}m{21mm}|c|c|c|c}
\toprule
\textbf{\diagbox[width=20mm]{Metric}{Model}}
 & \textbf{Tiny-WiFo} & \textbf{Version 1} & \textbf{Version 2} & \textbf{Version 3} \\ 
\midrule
Parameters (M)      & $5.5$   & $4.4$ & $1.4$ & $0.86$ \\ \midrule
Performance (dB) & \makecell{{$-6.75$}\\{$-7.92$}}  & \makecell{{$-6.21$}\\{$-7.13$}} & \makecell{{$-5.97$}\\{$-7.07$}}  & \makecell{{$-5.56$}\\{$-6.50$}}\\ 
\bottomrule
\end{tabular}}
\label{tradeoff}
\vspace{-0.7em}
\end{table}

To investigate the performance–size tradeoff, we further compress Tiny‑WiFo by reducing its encoder/decoder depth and hidden dimensions, obtaining three variants (Versions 1–3). Compared with teacher, Version 1 halves both depth and hidden dimension; Version 2 keeps the depth but reduces hidden dimension to one-quarter; Version 3 halves the depth and reduces hidden dimension to one-quarter. As shown in Table \ref{tradeoff}, parameter reduction leads to a graceful decline in NMSE. This demonstrates a flexible design space where users can select an appropriate model size according to their accuracy requirements and available computational resources.

\vspace{-0.9em}
\subsection{Scalability Evaluation: A Case Study on CSI Feedback}
Taking CSI feedback as an example, we evaluate the scalability of MCAKD scheme on other wireless FMs. We choose WiFo-CF \cite{wifo-cf} as the teacher and build a compact student model, termed Tiny-WiFo-CF. The student halves both depth and hidden dimension, reducing its parameter count to only $13.6\%$ of the teacher's. During CSI feedback, the user employs a non-uniform $\mu$-law quantizer with $3$-bit quantization bit-width, and the uplink SNR is fixed at 10 dB. WiFo-CF achieves an average NMSE of $-16.9$ dB on D1-D16 datasets, whereas the distilled student reaches $-15.4$ dB and the non-distilled model attains $-14.6$ dB. On D17 and D18 datasets, the distilled model achieves an average NMSE of $-14.02$ dB, approaching the teacher model's $-14.92$ dB and surpassing the non-distilled version by $1$ dB. These results demonstrate that MCAKD can be applied to other transformer-based wireless FMs, underscoring its generality and practical value.}

\vspace{-0.7em}
\subsection{Parameters and Inference Time}
\begin{table}[!t]
\setlength{\tabcolsep}{1.1pt}     
\renewcommand{\arraystretch}{0.6}
\centering
\caption{Network parameters and inference time.}
{\scriptsize
\begin{tabular}{>{\centering\arraybackslash}m{24mm}|c|c|c|c|c}
\toprule
\textbf{\diagbox[width=23mm]{Metric}{Model}}
 & \textbf{Tiny-WiFo} & \textbf{\makecell[c]{Teacher\\ WiFo}} & \textbf{Transformer} & \textbf{LSTM} & \textbf{LLM4CP*} \\ 
\midrule
Parameters (M)      & $5.5$   & $21.60$ & $0.91$ & $1.13$  & $83.32$ \\ \midrule
Inference Time (ms) & $7.04$  & $9.66$ & $6.24$ & $5.21$ & $7.34$ \\ 
\bottomrule
\end{tabular}}
\label{time}
\vspace{-1.6em}
\end{table}
In Table \ref{time}, we compare the parameter count and inference time for each model. The measurements are conducted on a workstation equipped with $4$ NVIDIA GeForce RTX 4090 GPUs, an AMD EPYC 7763 64-core CPU.
Table \ref{time} presents the parameter count of each model together with its inference time, measured on a batch size of $8$ with test samples taken from dataset D1. 
Compared to the teacher WiFo, Tiny-WiFo achieves nearly a $75$\% reduction in parameter count while also reducing the average inference latency by over $25$\%. The significant reduction in parameters enables Tiny-WiFo to be deployed on less powerful edge devices, as its inference and fine-tuning processes require considerably less GPU memory and power consumption. Although the Transformer-based and LSTM-based scheme have a small parameter count, they only offer slightly faster runtime but suffer from a significant  performance gap. Tiny-WiFo has over $15\times$ fewer parameters than LLM4CP*, while maintaining comparable inference time and achieving superior prediction performance, highlighting its exceptional parameter efficiency.

We also investigate whether Tiny-WiFo could achieve real-time inference on edge devices. We perform INT8 quantization on Tiny-WiFo and its teacher, and test their average inference time on the NVIDIA Jetson AGX Orin (a high-performance module for edge devices like drones). {\color{black}The teacher WiFo requires $2.3$ ms per inference, while Tiny-WiFo achieves only $1.6$ ms. This $0.7$ ms saving allows a larger portion of the upcoming data symbols to be demodulated with the updated CSI, thereby boosting the overall system throughput. }

\vspace{-0.8em}
\section{Conclusions}
This letter proposed a novel KD scheme termed MCAKD to achieve the compression of transformer-based wireless FMs.  MCAKD incorporated two critical modules to selectively identify essential knowledge and enhance training efficiency. By applying MCAKD to WiFo, we constructed a lightweight yet powerful channel prediction model called Tiny-WiFo.
Experimental results showed that Tiny-WiFo retained over $98\%$ of the teacher’s performance with only $5.5$M parameters and a $1.6$ ms inference time on the NVIDIA Jetson AGX Orin while maintaining excellent zero-shot generalization for practical deployment.

\newpage

 




\vfill

\end{document}